\documentclass[aps,prb,showpacs,twocolumn,superscriptaddress]{revtex4}

\usepackage[centerlast]{caption}
\usepackage{etex}
\usepackage{amsmath}
\usepackage{graphicx}
\usepackage{listings}
\usepackage{enumerate}
\usepackage{subfigure}
\usepackage{rotating}
\usepackage{textcomp}

\begin{document}

\title{General-purpose molecular dynamics simulations on GPU-based clusters}

\author{Christian R.\ Trott}
\email{christian.trott@tu-ilmenau.de}
\homepage{http://www.tu-ilmenau.de/theophys2}
\affiliation{Institut f\"ur Physik, University of Technology
  Ilmenau, 98684 Ilmenau, Germany}
\author{Lars Winterfeld}
\email{lars.winterfeld@tu-ilmenau.de}
\affiliation{Institut f\"ur Physik, University of Technology
  Ilmenau, 98684 Ilmenau, Germany}
\author{Paul S.\ Crozier}
\email{pscrozi@sandia.gov}
\affiliation{Scalable Algorithms, Sandia National Laboratories, P.O. Box 5800, MS 1322, Albuquerque, New Mexico, 87185-1322, USA}


\begin{abstract}
We present a GPU implementation of LAMMPS, a widely-used parallel molecular dynamics (MD) software package, and show 5x to 13x single node speedups versus the CPU-only version of LAMMPS.  
This new CUDA package for LAMMPS also enables multi-GPU simulation on hybrid heterogeneous clusters, using MPI for inter-node communication, CUDA kernels on the GPU for all methods working with particle data, and standard LAMMPS C++ code for CPU execution.
Cell and neighbor list approaches are compared for best performance on GPUs, with thread-per-atom and block-per-atom neighbor list variants showing best performance at low and high neighbor counts, respectively.
Computational performance results of GPU-enabled LAMMPS are presented for a variety of materials classes (e.g. biomolecules, polymers, metals, semiconductors), along with a speed comparison versus other available GPU-enabled MD software.
Finally, we show strong and weak scaling performance on a CPU/GPU cluster using up to 128 dual GPU nodes.

\end{abstract}
\maketitle

\section{Introduction}

During the last thirty years high performance computing (HPC) has become an increasingly-important tool in scientific research.
HPC studies enhance understanding of experimental findings, allow researchers to test theories on model systems, and even make it possible to investigate phenomena which cannot be investigated via classical experiments.
One class of computer experiments is of special interest: molecular dynamics (MD) simulations.
MD is used to simulate materials on an atomic (or coarser-grained) level using various interaction models.
Through advances in compute capabilities and algorithms, MD simulations have gradually expanded their range of applicability from modeling tiny systems of a few hundred atoms for up to a few thousand time steps, to performing short multi-billion atom simulations or multi-billion time-step simulations of smaller systems.
While this is already impressive in itself, a single cubic centimeter of matter contains on the order of $10^{23}$ atoms, and to model only one second of its time propagation, $10^{15}$ time steps (typically a femtosecond each) would be required.
Therefore, the interest in accelerating MD simulations is unstinting and of great interest for many computational scientists.

Easily programmable graphics cards (GPUs) represent a disruptive technology development that allows radical departure from recent years' gradual improvements in MD simulation speed. 
By harnessing the compute capability of GPUs, MD practitioners will be able to simulate much larger systems for much longer simulated times.
GPUs represent a jump in the performance-to-cost ratio of at least a factor of five.
GPUs also achieve more flops-per-watt than corresponding CPU hardware, making next-generation GPU-based HPC supercomputers more feasible from an operating energy cost perspective.
The CUDA programming language is currently the most widely used programming model for GPUs.
Since its introduction, many scientific programmers have used CUDA to write extremely fast software, thereby enabling previously-impossible investigations.
Among those are also a number of MD codes which have shown speed-ups of 5-100x over existing CPU-based codes.

In this paper, we present our own implementation of a GPU-MD code called LAMMPS$_{\rm CUDA}$, which is introduced as an extension to the widely used MD code LAMMPS\cite{LAMMPS}.
With its 26 different force fields, LAMMPS$_{\rm CUDA}$ can model atomic, polymeric, biological, metallic, granular, and coarse-grained systems up to 20 times faster than a modern quad core workstation by harnessing a modern GPU.
At the same time it offers unprecedented multi-GPU support for an MD code.
By providing very effective scaling of simulations on up to hundreds of GPUs, LAMMPS$_{\rm CUDA}$ enables scientists to harness the full power of the world's most advanced supercomputers, such as the world's fastest supercomputer, the Tianhe-1A at the Chinese National Supercomputing Center in Tianji\cite{top500Nov2010}.

We start with a description of the design objectives of our implementation and an overview of the features of LAMMPS$_{\rm CUDA}$.
Then we discuss aspects of our GPU implementations of LAMMPS's pair force calculations.
Performance results are then presented for various MD simulations on single GPUs.
This is followed by a discussion of strategies that enable GPU-based MD codes to scale well on systems with many GPUs. 
We also report and analyze LAMMPS$_{\rm CUDA}$ performance results on NCSA's Lincoln cluster, using up to 256 GPUs.

Parameters of the benchmark simulations are listed in Appendix \ref{sec:app_simulations} and hardware configuration are given in Appendix \ref{sec:app_hardware}.

\section{Design objectives, features and usage}
\label{sec:design}
Numerous GPU-MD codes have been under development during the past several years.
Some of those are new codes (HOOMD\cite{HOOMD}, AceMD\cite{ACEMD}), others are extensions or modifications of existing codes (e.g. NAMD\cite{NAMD}, Amber\cite{AMBER}, LAMMPS\cite{LAMMPS-GPU}).
Most of these projects are of limited scope and cannot compete with the rich feature sets of legacy CPU-based MD codes.
This is not surprising considering the amount of development time which has been spent on the existing codes; many of them have been under development for more than a decade.
Furthermore, some of these GPU-MD codes have been written to accelerate specific compute-intensive tasks, limiting the need to implement a broad feature set.
Our goal is to provide a GPU-MD code that can be used for simulation of a wide array of materials classes (e.g. glasses, semiconductors, metals, polymers, biomolecules, etc.) across a range of scales (atomistic, coarse-grained, mesoscopic, continuum).
LAMMPS can perform such simulations on CPU-based clusters.
It is a classical MD code that been under development since the mid 1990s, is freely-available, and includes a very rich feature set.
Since building such simulation software from scratch would be an enormous task, we instead leverage the tremendous effort that has gone into LAMMPS, and enable it to harness the compute power of GPUs.
We have written a LAMMPS "package" that can be built along with the existing LAMMPS software, thereby preserving LAMMPS' rich feature set for users while yielding tremendous computational speedups.
Other important LAMMPS features include an extensive scripting system for running simulations, and a simple-to-extend and modular code infrastructure that allows for easy integration of new features.
Most importantly it has an MPI-based parallelization infrastructure that exhibits good scaling behavior on up to thousands of nodes.
Finally, starting with an existing code like LAMMPS and building GPU versions of functions and classes one by one allows for easy code verification.

\noindent Our objectives can be summarized as follows (in order of decreasing priority):

\begin{enumerate}[(i)]
\item maintain the rich feature set and flexibility of LAMMPS,
\item achieve the highest possible speed-ups,
\item allow good parallel scalability on large GPU-based clusters,
\item minimize code changes,
\item write the code so that it is easy to maintain,
\item include GPU support for the full list of LAMMPS capabilities,
\item make the GPU capabilities easy for LAMMPS users to invoke.
\end{enumerate}

\noindent All of these design objectives have implications for design decisions, yet in many cases they are competing objectives.
For example objective (i) implies that the different operations of a simulation have to be done by different modules, and that the modules have to be able to be used in any combination requested by the user.
This in turn means that data, such as the particle positions, are loaded multiple times during a single simulation step from the device memory, which results in a considerably negative effect on the performance of the simulation. 
Another slight performance hit is caused by the use of templates for the implementation of pair forces and communication routines. 
While this greatly enhances maintainability, it adds some computational overhead. 
By keeping full compatibility with LAMMPS we were able to minimize the GPU-related changes that users will need to make to existing input scripts. 
In order to use LAMMPS$_{\rm CUDA}$ it is often enough to add the line ``accelerator cuda'' at the beginning of an existing input script. 
This triggers use of GPUs for all GPU-enabled features in LAMMPS$_{\rm CUDA}$, while falling back to the original CPU version for all others.

Another big influence on design decisions comes from the limiting factors of the targeted architecture.
Since those have been discussed in detail elsewhere\cite{CUDA-PG}, here we only list the most important factors:

\begin{enumerate}[(a)]
\item in order to use the full GPU, thousands or even tens of thousands of threads are needed,
\item data transfer between the host and the GPU is slow,
\item the ratio of device memory bandwidth to computational peak performance is much smaller than on a CPU,
\item latencies of the device memory are large,
\item random memory accesses on the GPU are serialized.
\item 32 threads are executed in parallel

\end{enumerate}

\begin{figure}[h!tb]
\begin{center}
  \includegraphics[width=0.25\textwidth]{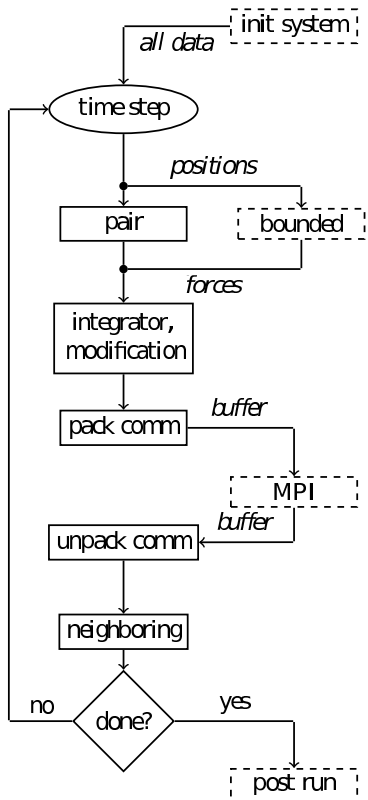}
\end{center}
            \vspace{-0.5cm}
  \caption{LAMMPS$_{\rm CUDA}$ work-flow, dashed boxes are done on the CPU, while solid boxes are done on the GPU.}\label{fig:workflow}
\end{figure}

Considering (b), we decided to minimize data transfers between device and host by running as many parts of the simulation as possible on the GPU.
This distinguishes our approach from other GPU extensions of existing MD codes, where only the most computationally-expensive pair forces are calculated on the GPU.
A work-flow chart of our implementation is shown in Figure \ref{fig:workflow}.

Currently LAMMPS$_{\rm CUDA}$ supports 26 pair force styles; long range coulomb interactions via a particle-particle/particle-mesh (PPPM) algorithm; NVE, NVT, and NPT integrators; and a number of LAMMPS ``fixes''.
In addition, pair force calculations on the GPU can be overlapped with bonded interactions and long range coulomb interactions if those are evaluated on the CPU.
All of the bond, angle, dihedral, and improper forces available in the main LAMMPS program can be used.
Simulations can be performed in single (32 bit floats) and double (64 bit floats) precision, as well as in a mixed precision mode, where only the force calculation is done in single precision while the time integration is done in double precision.
In addition to the requirements of LAMMPS, only the CUDA toolkit (available for free from NVIDIA) is needed.
Currently only NVIDIA GPUs with a compute capability of 1.3 or higher are supported.
This includes GeForce 285, Tesla C1060 as well as GTX480 and Fermi C2050 GPUs.
The package is available under the GNU Public License and can be downloaded from {\tt http://code.google.com/p/gpulammps/}, where detailed installation instructions and feature lists can be found.
LAMMPS$_{\rm CUDA}$, which is encapsulated in the USER-CUDA package of LAMMPS, should not be confused with LAMMPS' ``GPU'' package, which has some overlapping capabilities (see Figures \ref{fig:bench_system_size} and \ref{fig:scaling}) and is also available from the same website.

\section{Pair Forces}
We analyzed two variants of short range force calculations: a cell list approach and a neighbor list approach.
While most CPU-based MD codes use a neighbor list approach for the force calculation, it has been suggested\cite{HarvestingCellLists} that the cell list approach is better suited for GPU implementations.

\subsection{Cell list approach}
The idea of the cell list approach is a spatial decomposition of the simulation box into a regular grid of small sub-cells, with a maximum number of atoms per cell.
Because LAMMPS uses neighbor lists, additional effort is required to re-order the existing data structures for the GPU calculation and to convert the data back into the original LAMMPS format for every usual computation not done on the GPU.

In order to implement this idea on the GPU, we associate every \textit{cell} with a CUDA thread \textit{block} and have each of the $n_{cell\_nmax}$ \textit{threads} of it calculate the forces for one \textit{particle} in the cell.

\begin{figure}[h!]
	\caption{Cell list approach}
	\subfigure[ 2D depiction of cell lists]
	{
		\label{fig:part_cell}
		\includegraphics[width=0.2\textwidth]{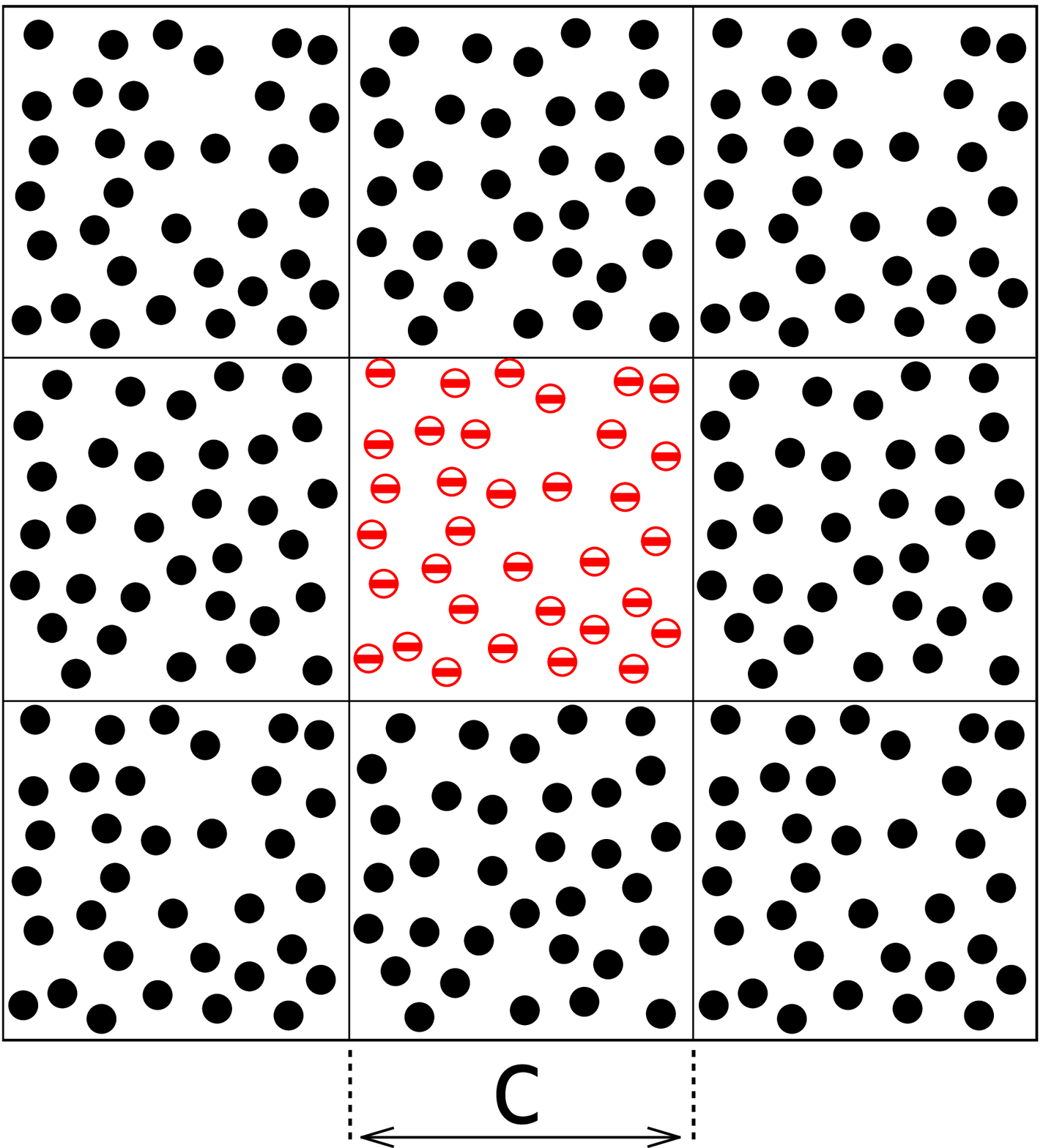}
	}
	\subfigure[ cell update pattern]
	{
		\label{fig:cell_neighbor_pattern}
		\includegraphics[width=0.2\textwidth]{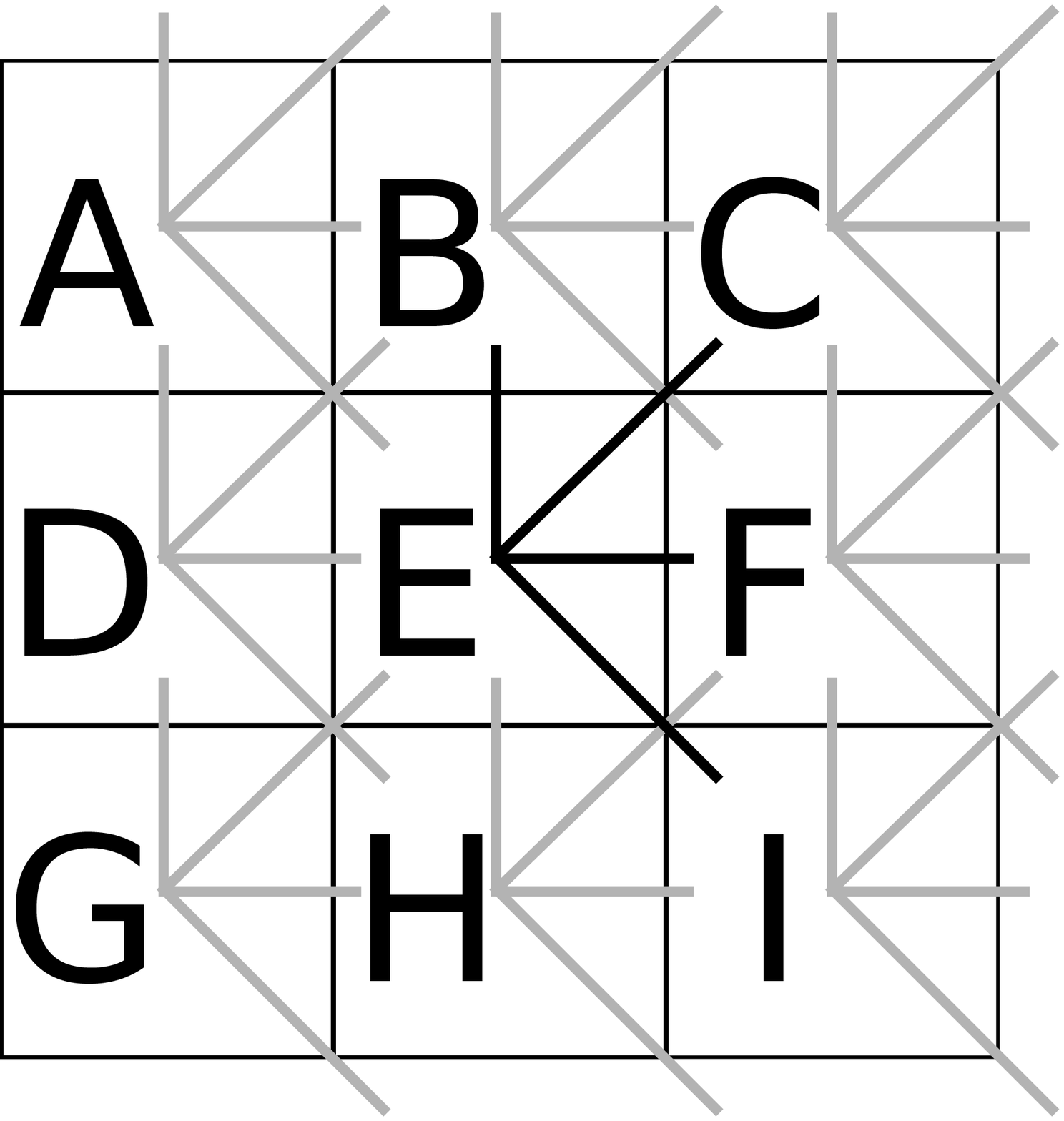}
	}
	\label{fig:cell_lists}
\end{figure}

Furthermore it is necessary to choose the cell size $c$ (see Fig.
\ref{fig:cell_lists}(a)) and the maximum number of atoms per cell $n_{cell\_nmax}$.
For a given force cut-off radius $r_c$, we choose $c \approx 2 \: r_c$ in order to keep the average distance between particles in the cell $\approx r_c$ and to limit the frequency of re-assigning atoms to their cells.
Also, $c$ should be large enough to contain at least 32 particles in order to not to leave GPU threads idle.
Accordingly, $n_{cell\_nmax}$ is automatically chosen as a multiple of 32, depending on the particle density.

When performing the force calculations in the cell list approach, at least two more optimizations can be used.
The first is to use Newton's third law $\vec F_{ab} = - \vec F_{ba}$ to save half of the force calculation time.
In 2D, forces need to be explicitly computed for only 4 of 8 the neighboring cells, with the other 4 obtained via Newton's third law during other cells' updates.
Figure \ref{fig:cell_lists}(b) depicts an example of such an update pattern, with the explicitly-computed neighbors of cell E connected with cell E by a solid black line, and the other 4 neighbors of cell E connected with cell E by solid gray lines.
Every cell then follows this pattern, and the interactions between all neighboring cells are then considered exactly once, as verified for cell E.
In 3D, only 13 of the 26 neighboring cells are explicitly considered.
Note, however, that not every selection of 13 neighboring cells fulfills the required periodicity.

Execution of GPU thread blocks can be in any order, whether in sequence or in parallel.
Therefore, write conflicts may occur.
For example, in Figure \ref{fig:cell_lists}(b), cell A and cell D might try to update the forces in cell B at the same time.
In order to avoid such a write conflict and a resulting error in the calculation, the code has been written to execute only non-interfering groups of cells simultaneously.
If only one neighbor shell needs to be considered, there are six such groups in 2D and 18 such groups in 3D.
This does not significantly affect performance since $N$ groups are executed, each in approximately $\frac{1}{N}$ of the original time.

The second optimization is the use of shared memory for the positions of the particles in the neighboring cells.
If a cell contains more atoms than will fit in shared memory, the particles have to be loaded to shared memory in groups one after another. 
For a more detailed discussion on this topic, see \cite{LarsBA}.

\subsection{Neighbor list approach}
In designing a neighbor list approach that uses blocks of threads, it becomes clear that there are two main ways that the force calculation work can be divvied up among the threads.
The first possibility is to use one thread per atom (TpA), where the thread loops over all of the neighbors of the given atom.
The second possibility is to use one block per atom (BpA), where each of the threads in the block loop over its designated portion of the neighbors of the given atom.
In the following, pseudo-code for both algorithms are given:\newline\newline

\noindent TpA algorithm:
\begin{lstlisting}
1  i = blockId*ThreadsPerBlock+threadId;
2  load(i) // coalesced access
3  for(jj = 0; jj<numneigh[i]; 
4      jj++) {
5   j <- neighbors[i][jj]
6   load(j) // random access
7   ftmp+=calcPairForce(i,j)
8  }
9 
10 ftmp -> f[i] // coalesced access
\end{lstlisting}

\noindent BpA algorithm:
\begin{lstlisting}
1  i = blockId
2  load(i) // coalesced access
3  for(jj = 0; jj<numneigh[i]; 
4      jj+=ThreadsPerBlock) {
5   j <- neighbors[i][jj]
6   load(j) // random access
7   ftmp+=calcPairForce(i,j)
8  }
9  reduce(ftmp)
10 ftmp -> f[i] // coalesced access
\end{lstlisting}

Both algorithms ostensibly have the same number of instructions; however, when considering looping it becomes clear that the BpA algorithm requires the execution of a larger total number of lines of code.
The BpA algorithm also requires the relatively expensive reduction of $ftmp$ that is not required by the TpA algorithm.
BpA also requires use of a much larger total number of blocks.
For further clarification, Table \ref{tab:BpA-TpA-NExec} lists the number of times each line of code is executed, taking into account the number of blocks used, and considering that 32 threads of each block are executed in parallel.

\begin{table}
\caption{\label{tab:bf} Number of executions per line for the BpA and TpA algorithms}
\label{tab:BpA-TpA-NExec}
\begin{center}
\begin{tabular}{@{\hspace{1em}}c@{\hspace{1em}}|@{\hspace{1em}}c@{\hspace{1em}}c@{\hspace{1em}}}
\hline\hline
Lines  & TpA & BpA \\
1,2,9,10:	& 	natoms/32 		&	natoms\\
5,6,7: 		&	(natoms/32)*nneigh 	&	natoms*(nneigh/32)\\
\hline\hline
\end{tabular}
\end{center}
\end{table}

While this seems to indicate that TpA would always be faster, one has to take into account cache usage as well.
In order to reduce random accesses in the device memory while loading the neighbor atoms (limiting factor (e)), one can cache the positions using the texture cache.
(We also tested global cache on Fermi GPUs, but it turns out to be slower due to its cache line size of 128 bytes.)
This strategy improves the speed of both algorithms considerably, but it helps BpA more than TpA.
The underlying reason is that less atoms are needed simultaneously with BpA than with TpA.
As a result, BpA allows for better memory locality, and therefore the re-usage of data in the cache is increased (assuming atoms are spatially ordered).
Revisiting table \ref{tab:BpA-TpA-NExec} makes it evident that this better cache usage becomes increasingly important with an increasing number of neighbors, corresponding to an increased pair cutoff distance.

\begin{figure}[t]
	\includegraphics[width=0.45\textwidth, ]{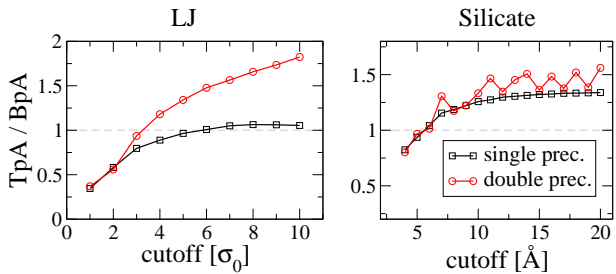}
	\caption{Computation time comparison of BpA and TpA algorithms for two different benchmark systems in single and double precision.
The ratio of the force calculation time using the TpA algorithm and the force calculation time using the BpA algorithm is shown as a function of the cutoff.
System: LJ (32k atoms), Silicate (12k atoms) (see Appendix \ref{sec:app_simulations}); Hardware: CL (see Appendix \ref{sec:app_hardware})}
	\label{fig:bench_BpAvsTpA}
\end{figure}

Consequently, one can expect a crossover cutoff for each type of pair force interaction, where TpA is faster for smaller cutoffs and BpA for larger.
Unfortunately it is hard to predict where this cutoff lies.
It not only depends on the complexity of the given pair force interaction, but also on the hardware architecture itself.
Therefore, a short test is the best way to determine the crossover cutoff.
The timing ratios shown in Figure \ref{fig:bench_BpAvsTpA} indicate that the force calculation time can depend significantly on the use of BpA or TpA.
Generally the differences are larger when running in single precision than when running in double precision.
For the LJ system, an increase of the cutoff from $2.5\sigma_0$ to $5.0\sigma_0$ can turn the 30\% TpA advantage into a 30\% BpA advantage.
Therefore we decided to implement both algorithms, and allow dynamic selection of the faster algorithm using a built-in mini benchmark during the setup of the simulation.
This ensures that the best possible performance is achieved over a wide range of cutoffs.
While our particular findings are true only for NVIDIA GPUs, one can expect that similar results would be found on other highly parallel architectures with comparable ratios of cache to computational power.

We tested the cell list approach and the neighbor list approach for a small LJ system as a function of cutoff radius (see Fig. \ref{fig:cell_neigh}).
For the neighbor list approach, three distinct regions can be seen, as labeled in the Figure.
For regions Ia and Ib the TpA method is faster than the BpA method.
Above $r_c=3.5\sigma$ the BpA algorithm is faster than the TpA method, so the code automatically switches to BpA, resulting in a different slope for region II.
The two different slopes in Ia and Ib are most likely a result of the limited texture cache size.
For small cutoffs, most neighbors fit into the texture cache, facilitating efficient re-usage of data.
But at some point the collective number of neighbors becomes large enough that the texture cache can no longer be used efficiently.
This changes the scaling behavior as a function of the cutoff radius.
 
\subsection{Comparing the cell and neighbor list approaches}
\label{sec:comp_neigh_cell}
\begin{figure}[t]
	\includegraphics[width=0.45\textwidth]{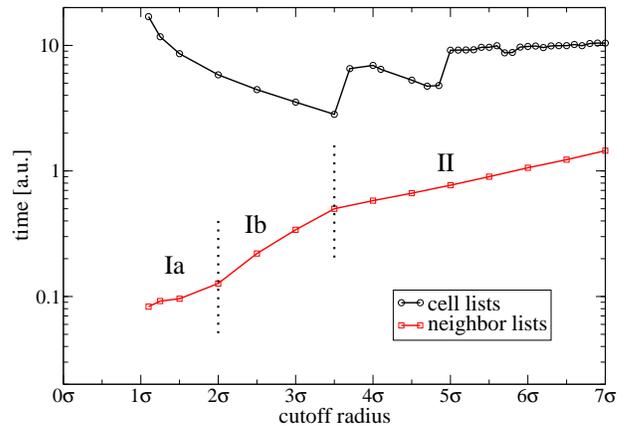}
	\caption{ Computation time comparison of cell-list-based and neighbor-list-based force calculations for a small LJ system, $n \approx 20,000$ particles.
In region I ($r_c<3.5\sigma$) the TpA algorithm is used, while in region II the BpA algorithm is used.
(The faster algorithm is automatically selected for each region.)
In sub-region Ia texture cache is used effectively by the TpA algorithm, but in region Ib the cache must be flushed frequently.
The jumps in the cell list curve are caused by the GPU requirement of $n_{cell\_nmax}$ being a multiple of 32 and the resulting unsteady proportion of started threads versus those that are actually needed. System: LJ (see Appendix \ref{sec:app_simulations}); Hardware: WS$_A$ (see Appendix \ref{sec:app_hardware})}
	\label{fig:cell_neigh}
\end{figure}

Figure \ref{fig:cell_neigh} clearly demonstrates that the cell-list-based force evaluation is considerably slower than the neighbor-list-based approach for all cut-offs.
In this section, we will make a simple argument why this is not only true for the above example, but has to be expected in general.

Figure \ref{fig:cell_neigh} is based on an earlier program version that still featured both cell and neighbor lists.
Due to the weak performance of the cell list approach, we have completely dropped it and have focused our efforts on the optimization of the neighbor-list-based force calculation.
While further improvements might have been possible for the cell list approach as well, the superiority of the neighbor list approach appears to be inevitable, as explained below.

Obviously, the time $t$ for processing a single interaction force consists of two parts: memory access (i.e. reading the other atom's position) and the evaluation of the force formula.
Na\"{\i}vely one might assume that the total time $T$ needed for all force calculations equals $t$ times the number of interactions.
However, both the cell and the neighbor list algorithms first load all potential interaction partners to check if they are within the cut-off radius $r_c$.
Whenever one thread finds an atom close enough ($r \leq r_c$) and evaluates the force formula, the other threads processing interactions with $r > r_c$ have to wait until every thread in the warp has completed its calculations.
Therefore, the time for both memory access and for the evaluation of the force formula scale with the number of possible interaction partners $N$, i.e. it is reasonable to say $T = t \cdot N$.
Still both factors depend on which algorithm is chosen (cell or neighbor list).
To determine which is faster, we examine the ratio of their computational times:
\begin{equation}
 \frac { T_{\text{cell}} } { T_{\text{neigh}} }
 =
 \frac { t_{\text{cell}} } { t_{\text{neigh}} }
  \cdot
 \frac  { N_{\text{cell}} } { N_{\text{neigh}} } \; .
\end{equation}

\begin{figure}[h!]
	\subfigure[]
	{
		\label{fig:particles_cell_size_a}
		\includegraphics[width=0.2\textwidth]{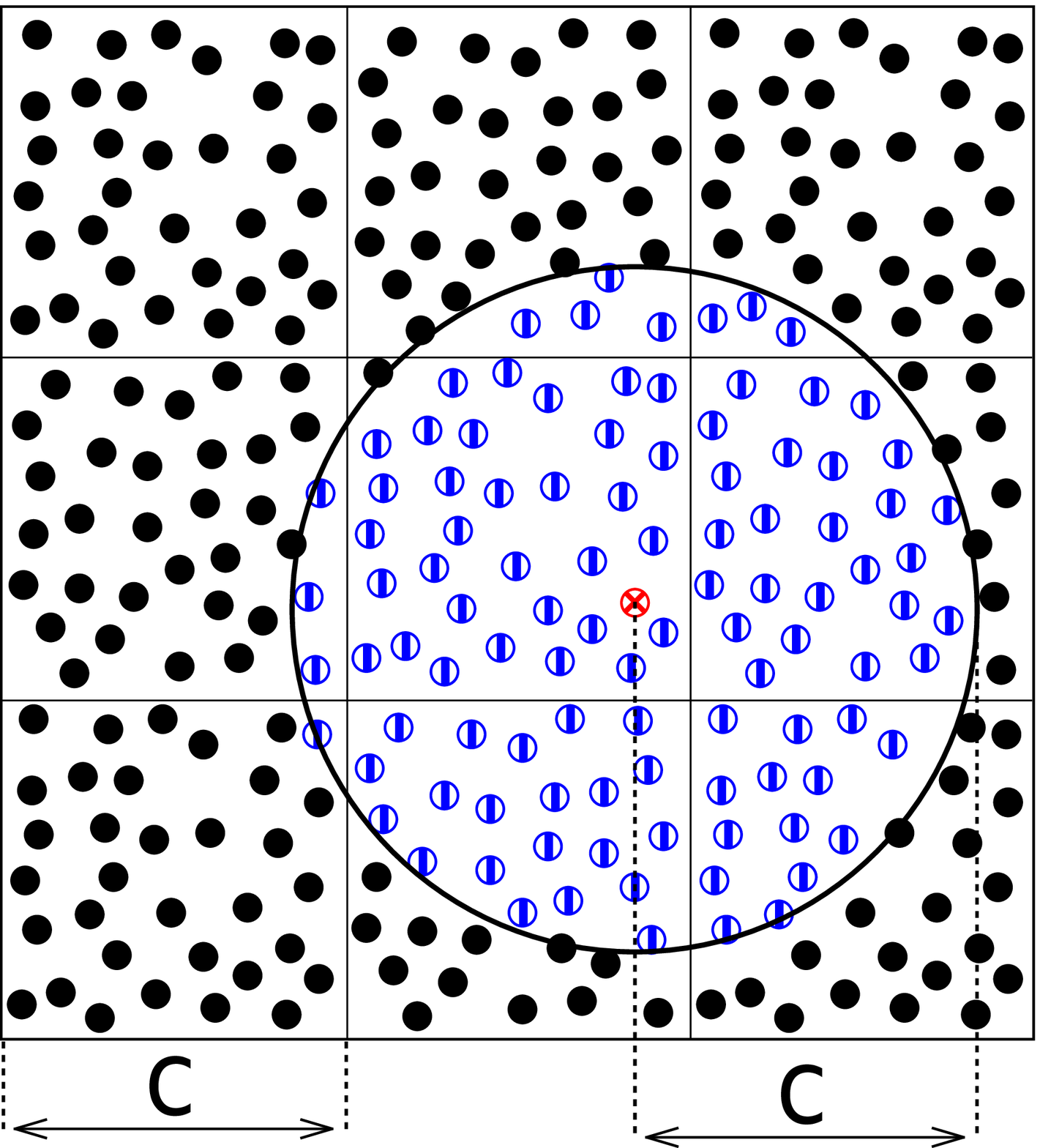}
	}
	\subfigure[]
	{
		\label{fig:alphagamma}
		\includegraphics[width=0.23\textwidth, clip=]{celllist-alpha-gamma}
	}
	\caption{(a) Required memory elements, depicted in 2D \;\; (b) $\gamma$ is the fraction of the total force calculation time spend on memory accesses.
$\alpha$ is the factor by which coalesced memory accesses are faster than random memory accesses.
Along the solid line the cell-list-based force calculation is as fast as a neighbor-list-based pair force calculation.
The dotted lines are the asymptotic limits ($\gamma \approx 70\%$, $\alpha \approx 3.3$).}
	\label{fig:cell_benchmark_explanation}
\end{figure}

For geometric reasons, $\frac  { N_{\text{cell}} } { N_{\text{neigh}} } = \frac{14 \cdot 3}{4 \pi} \approx 3.3 > 1$.
Figure \ref{fig:cell_benchmark_explanation}(a) illustrates the 3D situation in a 2D sketch.
The cell list approach requires loading all atom positions from $\frac{3^3 - 1}{2} + 1 = 14$ surrounding (cubic) cells, each of edge length $c = 2 r_c$, while the neighbor list includes only atoms within a sphere of radius $c$. 
The cell list approach is wasteful in the sense that many non-interacting atoms are loaded into memory.
Since the cell list approach requires the loading of roughly 3.3 times more data into memory than the neighbor list approach, and since the time for the evaluation of the force formula is the same in both cases, the cell list approach can only be faster if its memory access time is smaller.
This could be possible due to coalesced memory accesses that can be done in the cell list approach.
In order to find out whether this is realistic, we model the situation with two parameters:
\begin{itemize}
 \item $\alpha$: the factor by which the coalesced memory accesses are faster than random accesses ($\alpha > 1$).
 \item $\gamma$: the fraction of $t$ which is assumed to be spent on memory accesses ($0 \leq \gamma \leq 1$).
\end{itemize}

Clearly, the cell lists need both high $\alpha$ and high $\gamma$ in order to gain the advantage with their faster memory accesses.
With a little algebra (see appendix \ref{sec:calc_comp}), we can quantify some limits for $\alpha$ and $\gamma$.
In order to make the cell list method viable, its memory accesses have to be at least 3.3 times faster than the memory accesses used in the neighbor list method, and at least 70\% of the total neighbor list force calculation time has to be spent on memory accesses.
In Figure \ref{fig:alphagamma} $\alpha(\gamma)$ is shown for $T_{\text{neigh}} = T_{\text{cell}}$.
While $\gamma > 70$\% is not unrealistic for computation of inexpensive pair forces, the use of texture reads in the pair force kernels limits the advantage of coalesced memory accesses considerably (i.e. decreasing $\alpha$), thus making the cell list approach always slower than the neighbor list approach.
In practice, the product of $\alpha$ and $\gamma$ is always below the solid line in Figure \ref{fig:alphagamma}, making the neighbor list approach the preferred alternative.

\section{Single node performance}

\begin{figure*}[t]
	\includegraphics[width=0.85\textwidth, ]{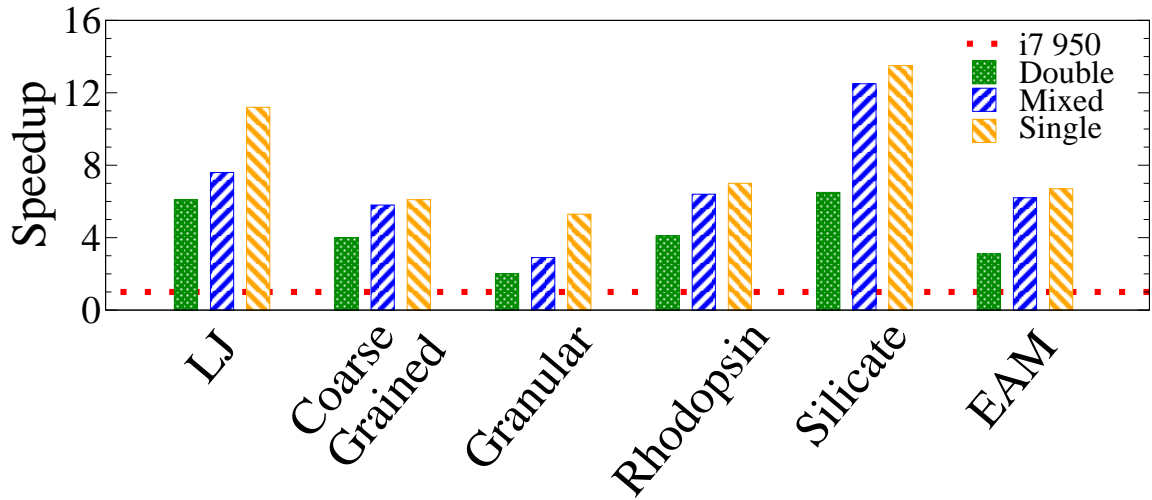}
	\caption{Typical speed-up when using a single GTX 470 GPU versus a Quad-Core Intel i7 950 for various system classes.
Systems: see Figure (see Appendix \ref{sec:app_simulations}); Hardware WS$_B$ (see Appendix \ref{sec:app_hardware})}
	\label{fig:bench_single_gpu}
\end{figure*}

\begin{figure}[tb]
  \includegraphics[width=0.45\textwidth]{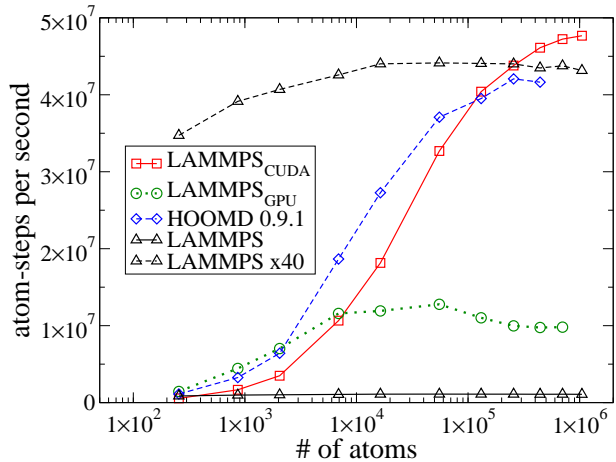}
  \caption{Performance in number of atom-steps per second of a GTX 470 GPU and a single core of an i7 950 CPU.
LAMMPS$_{\rm CUDA}$ approaches its maximum performance only for system sizes larger than 200,000 particles.
The same system was also run using HOOMD version 0.9.1 and the ``GPU'' package of LAMMPS. 
The CPU curve has also been plotted with a scaling factor of 40 to make it easier to see.
System: LJ (see Appendix \ref{sec:app_simulations}); Hardware: WS$_B$ (see Appendix \ref{sec:app_hardware})}\label{fig:bench_system_size}
\end{figure}

To assess the possible performance gains of harnessing GPUs, we have performed benchmark simulations of several important classes of materials.
Both the regular CPU version of LAMMPS and LAMMPS$_{\rm CUDA}$ were run on our workstation B (WS$_B$) with an Intel i7 950 quad core processor and a GTX 470 GPU from NVIDIA.
Simulations on the GPU were carried out in single, double, and mixed precision.
We compare the loop times for 10,000 simulation steps.
The results shown in Figure \ref{fig:bench_single_gpu} are proof of an impressive performance gain.
Even in the worst-case scenario, a granular simulation (which has extremely few interactions per particle), the GPU is 5.3 times as fast as the quad core CPU when using single precision and 2.0 times as fast in double precision.
In the best-case scenario the speed-up reaches a factor of 13.5 for the single precision simulation of a silicate glass involving long range coulomb interactions.
Single precision calculations are typically twice as fast as double precision calculations, while mixed precision is somewhere in between.
It is worthy to note that this factor of two between single and double precision is reached on consumer grade GeForce GPUs, despite the fact that their double precision peak performance is only 1/8th of their single precision peak performance.
This is a strong sign that LAMMPS$_{\rm CUDA}$ is memory bound.

Generally, the speed-up increases with the complexity of the interaction potential and the number of interactions per particle.
Additionally the speed-up also depends on the system size.
As stated in section \ref{sec:design} the GPU needs many threads in order to be fully utilized.
This means that the GPU cannot reach its maximum performance when there are relatively few particle-particle interactions.
This point is illustrated in Figure \ref{fig:bench_system_size}, where the number of atom-steps per second is plotted as a function of the system size.
As can be seen, at least 200,000 particles are needed to fully utilize the GPU for this Lennard-Jones system.
In contrast the CPU core is already nearly saturated with only 1,000 particles.
All systems used to produce Figure \ref{fig:bench_single_gpu} were large enough to saturate the GPU.

We have also plotted the performance curves for the GPU-MD program HOOMD (version 0.9.1) and the ``GPU'' package of LAMMPS in Figure \ref{fig:bench_system_size} for comparison purposes.
The characteristics of HOOMD are very similar to LAMMPS$_{\rm CUDA}$.
It reaches its top performance at about 200,000 particles.
Interestingly HOOMD is somewhat slower than LAMMPS$_{\rm CUDA}$ at very high particle counts, while it is significantly faster at system sizes of 16,000 particles and below.
This can probably be explained by the fact that HOOMD is a single GPU code, whereas LAMMPS$_{\rm CUDA}$ has some overhead due to its multi-GPU capabilities.
The ``GPU'' package of LAMMPS reaches its maximum performance at about 8,000 particles.
While it is faster than LAMMPS$_{\rm CUDA}$ for smaller systems (and even faster than HOOMD for fewer than 2,000 particles), it is significantly slower than LAMMPS$_{\rm CUDA}$ and HOOMD for this LJ system at large system sizes.
The reason is most likely that the ``GPU'' package of LAMMPS only off-loads the pair force calculations and the neighbor list creation to the GPU, while the rest of the calculation (e.g. communication, time integration, thermostats) is performed on the CPU.
This requires a lot of data transfers over the PCI bus, which reduces overall performance and sets an upper limit on the speed-up.
On the other hand, at very low particle counts the CPU is very efficient at doing these tasks that are less computationally demanding and memory bandwidth limited.
While a GPU has a much higher bandwidth to the device memory than does the CPU to the RAM, the whole data set can fit into the cache of the CPU for small system sizes. 
So for the smallest system sizes, the CPU can handle these tasks more efficiently than the GPU, leading to the higher performance of the ``GPU'' package for small system sizes.

\section{Scaling}
\begin{table*}
\begin{tabular}{p{0.08\textwidth}p{0.45\textwidth}p{0.45\textwidth}}
&
\begin{center}{\large \bf LJ}\end{center}&
\begin{center}{\large \bf Silicate (cutoff)}\end{center}\\
\begin{center}\begin{rotate}{90} \hspace{1cm}{\large \bf \hspace{0.5cm} Weak Scaling} \end{rotate}\end{center}
&
	\includegraphics[width=0.40\textwidth, ]{lj-melt-fixed-per-node3}
&
	\includegraphics[width=0.40\textwidth, ]{silicate-cut-fixed-per-node}
\\
\begin{center}\begin{rotate}{90} \hspace{1cm}{\large \bf \hspace{0.5cm} Strong Scaling} \end{rotate}\end{center}
&
	\includegraphics[width=0.40\textwidth, ]{lj-melt-fixed-size3}
&
	\includegraphics[width=0.40\textwidth, ]{silicate-cut-fixed-size}
\\
\end{tabular}

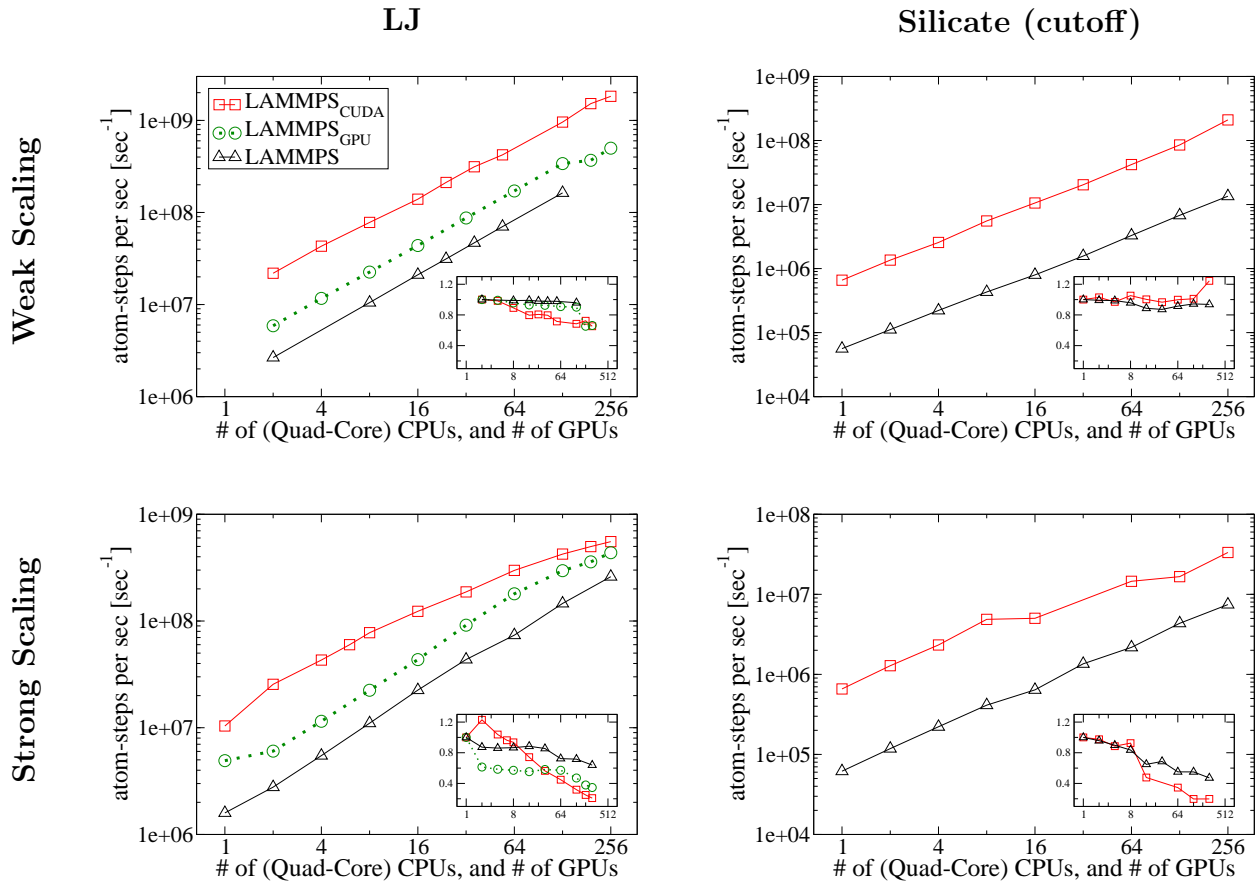
\captionof{figure}{Multi-node scaling comparison for a fixed system size setup (strong scaling), and a constant number of atoms per node setup (weak scaling).
High parallel efficiency is harder to achieve for the strong scaling case. The insets show the parallel efficencies. Each node includes 2 GPUs and 2 Quad-Core CPUs.
Systems: LJ, silicate (cutoff) (see Appendix \ref{sec:app_simulations}); Hardware: Lincoln (see Appendix \ref{sec:app_hardware})}
\label{fig:scaling}
\end{table*}

In order to simulate large systems within a reasonable wall clock time, modern MD codes allow parallelization over multiple CPUs.
LAMMPS's spatial decomposition strategy was specifically chosen to enable this parallelization, allowing LAMMPS to run efficiently on modern HPC hardware.
Depending on the simulated system, it has been shown to have parallel efficiencies \footnote{We define atom-steps per second as number of simulated steps $s$ times the number of atoms $n$ divided by the wall clock time $t$: $k=s \cdot n \cdot t^{-1}$.
Let $k_s$ denote the atom-steps per second for a single CPU run, and let $k_m$ denote the atom-steps per second for an $m$ CPUs run.
Parallel efficiency, $p$, is then the ratio of $k_s$ to $k_m$ multiplied by $m$: $p=\frac{k_s}{k_m}m$.} 
of 70\% to 95\% for up to several ten thousand CPU cores.
To split the work between the available CPUs, LAMMPS's spatial decomposition algorithm evenly divides the simulation box into as many sub-boxes as there are processors.
MPI is used for communication between processors.
During the run, each processor packs particle data into buffers for those particles that are within the interaction range of neighboring sub-boxes.
Each buffer is sent to the processor associated with each neighboring sub-box, while the corresponding data buffers from other processors are received and unpacked.

While the execution time of most parts of the simulation should in principle scale very well with the number of processors, communication time is a major exception.
With an increasing number of processors, the fraction of the total simulation time which is used for inter-processor communication increases.
This is already bad enough for CPU-based codes, where switching from 8 to 128 processors typically doubles or triples the relative portion of communication. 
But for GPU-based codes, the situation is even worse since the compute-intensive parts of the simulation are executed much faster (typically by a factor of 20 to 50 times).
It is therefore understandable why it is essential to perform as much of the simulation as possible on the GPU.
Consider the following example: in a given CPU simulation, 90\% of the simulation time is spent on computing particle interaction forces.
Running only that part of the calculation on the GPU, and assuming a 20-fold speed-up in computing the forces, the overall speed-up is only a factor of 6.9.
If we then assume that with an increasing number of processors the fraction of the force calculation time drops to 85\% in the CPU version, then the overall speed-up would be only a factor of 5.2.
On top of the usual parallel efficiency loss of the CPU code, additional parallel efficiency is lost for the GPU-based code if only calculating the pair forces on the GPU.

If one processes the rest of the simulation on the GPU as well, the picture gets somewhat better.
Most of the other parts of the simulation are bandwidth bound, i.e. typical speed-ups are around 5.
Taking the same numbers as before yields an overall speed-up of 15.4 and 13.8, respectively.
So if parts of the code that are less optimal for the GPU are also ported, not only will single node performance be better, but the code should also scale much better.
While the above numbers are somewhat arbitrary, they illustrate the general trend.

In order to minimize the processing time on the host, as well as minimize the amount of data sent over the PCI bus, LAMMPS$_{\rm CUDA}$ builds the communication buffers on the GPU.
The buffers are then transferred back to the host and sent to the other processors via MPI.
Similarly, received data packages are transferred to the GPU and only opened there.

Actual measurements have been performed on NCSA's Lincoln cluster, where up to 256 GPUs on 128 nodes were used (see Figure \ref{fig:scaling}).
We compare weak and strong scaling behavior of LAMMPS$_{\rm CUDA}$ versus the CPU version of LAMMPS for two systems: LJ and silicate (cutoff).
In the weak scaling benchmark, the number of atoms per node is kept fixed, such that the system size grows with increasing number of nodes.
In this way, the approximate communication-to-calculation ratio should remain fairly constant, and the GPUs avoid underutilization issues.
In the strong scaling benchmark, the total number of atoms is kept fixed regardless of the number of nodes used.
This is done in order to see how much a given fixed-size problem can be accelerated.
Note that in Figure \ref{fig:scaling}, we plot the number of quad-core CPUs rather than the number of individual cores.
Please also note that in general the Lincoln-cluster would not be considered a GPU-``based'' cluster since the number of GPUs per node is relatively small and two GPUs share a single PCIe2.0 8x connection.
This latter issue represents a potential communication bottleneck since there are synchronization points in the code prior to data exchanges.
Consequently, both GPUs on a node attempt to transfer their buffers at the same time through the same PCIe connection.
On systems where each of the (up to four) GPUs of a node has its own dedicated PCIe2.0 16x slot, the required transfer time would be as little as one fourth of the time on Lincoln, thus allowing for even better scaling.
Since Lincoln is not intended for large-scale simulations, it features only a single data rate (SDR) InfiniBand connection with a network bandwidth that can become saturated when running very large simulations.

Nevertheless, Figure \ref{fig:scaling} shows that very good scaling is achieved on Lincoln.
There, the number of atom-steps per second (calculated by multiplying the number of atoms in the system by the number of executed time-steps, and dividing by the total execution time) is plotted against the number of GPUs and quad-core CPUs that were used.
We tested two different systems: a standard Lennard-Jones system (density 0.84~$\sigma^{-3}$, cutoff 3.0~$\sigma_0$), and a silicate system that uses the Buckingham potential and cutoff coulombic interactions (density 0.09~{\AA}$^{-3}$, cutoff 15~{\AA}). 
While keeping the number of atoms per node constant, the scaling efficiency of LAMMPS$_{\rm CUDA}$ is comparable to that of regular CPU-based LAMMPS.
Even at 256 GPUs (128 nodes), a 65~\% scaling efficiency is achieved for the Lennard-Jones system that includes 500,000 atoms per node.
And a surprising 103~\% scaling efficiency is achieved for the silicate system run on 128 GPUs (64 nodes) and 34,992 atoms per node. 
This means that for the silicate system, 128 GPUs achieved more than 65 times as many atom-steps per second than 2 GPUs.
In this case, a measured parallel efficiency slightly greater than unity is probably due to non-uniformities in the timing statistics caused by other jobs running on Lincoln at the same time.

We were also able to run this Lennard-Jones system with LAMMPS's ``GPU'' package. 
As already seen in the single GPU performance, the GPU package is about a factor of three slower than LAMMPS$_{\rm CUDA}$ for this system.
The poorer single GPU performance leads to slightly better scaling for LAMMPS's GPU package.

Comparing the absolute performance of LAMMPS$_{\rm CUDA}$ with LAMMPS at 64 nodes gives a speed-up of 6 for the Lennard-Jones system and a speed-up of 14.75 for the silicate system. 
Translating that to a comparison of GPUs versus single CPU cores means speed-ups of 24 and 59, respectively.

Such larger speed-ups are observed up to approximately 8 nodes (16 GPUs) in the strong scaling scenario, where we ran fixed-size problems of 2,048,000 Lennard-Jones atoms and 139,968 silicate atoms on an increasing number of nodes. 
With 32 GPUs (16 nodes) the number of atoms per GPU gets so small (64,000 and 4,374 atoms, respectively) that the GPUs begin to be underutilized, leading to much lower parallel efficiencies (see Figure \ref{fig:bench_system_size}). 
At the same time, the amount of MPI communication grows significantly.
In fact, for the silicate system with its large 15~{\AA} cutoff, each GPU starts to request not only the positions of atoms in neighboring sub-boxes, but also positions of atoms in next-nearest neighbor sub-boxes.
This explains the sharp drop in parallel efficency seen at 32 GPUs. 
In consequence, 256 GPUs cannot simulate the fixed-size silicate system significantly faster than 16 GPUs. 
On the other hand, those 16 GPUs on 8 nodes are faster than all 1024 cores of 128 nodes when using the regular CPU version of LAMMPS. 

We also tested the ``GPU'' package of LAMMPS for strong scaling on the Lennard-Jones system. For this test, its parallel efficency is lower than that of LAMMPS$_{\rm CUDA}$ up to 32 GPUs. 
For more than 32 GPUs, the ``GPU'' package shows stronger scaling than LAMMPS$_{\rm CUDA}$. 
This can be ascribed to LAMMPS$_{\rm CUDA}$'s faster single node computations and subsequently higher communication-to-computation ratio.
(Note that each of the versions of LAMMPS discussed here have the same MPI communication costs.)
In LAMMPS$_{\rm CUDA}$, the time for the MPI data transfers actually reaches 50~\% of the total runtime when using 256 GPUs.

A simple consideration explains why the MPI transfers are a main obstacle for better scaling.
Since the actual transfer of data cannot be accelerated using GPUs, it constitutes the same absolute overhead as with the CPU version LAMMPS.
Considering that the rest of the code runs 15 to 60 times faster on a process-by-process basis, it is obvious that if 1~\% to 5~\% of the total time is spent on MPI transfers in the CPU LAMMPS code, communication can become the dominating time factor when using the same number of GPUs with LAMMPS$_{\rm CUDA}$.

\begin{figure}[tb]
  \includegraphics[width=0.45\textwidth]{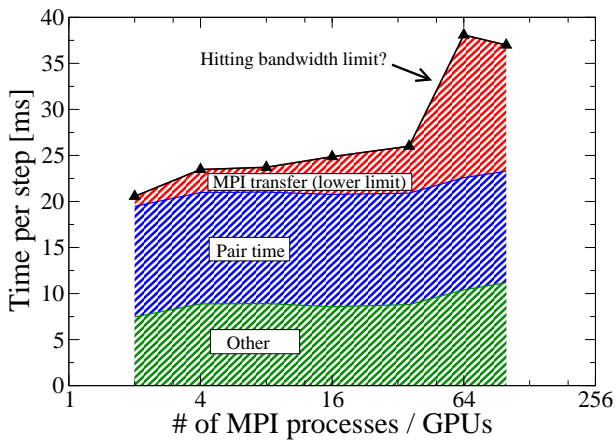}
  \caption{Portion of runtime spent on MPI transfers, pair force calculations, and other computations in a weak scaling benchmark. The increase of the total runtime is caused almost completely by the increase in the MPI transfer times.  System: LJ (see Appendix \ref{sec:app_simulations}); Hardware: Lincoln (see Appendix \ref{sec:app_hardware})}\label{fig:bench_mpi}
\end{figure}

That the MPI transfer time is indeed the main cause of the poor weak scaling performance can be shown by profiling the code.
Figure \ref{fig:bench_mpi} shows the total simulation time of the Lennard-Jones system versus the number of GPUs used.
It is broken down into the time needed for the pair force calculation, a lower estimate of the MPI transfer times and the rest.
The lower estimate of the MPI transfer time does not include any GPU$\leftrightarrow$host communication.
It only consists of the time needed to perform the MPI send and receive operations while updating the positions of atoms residing in neighboring sub-boxes.
All other MPI communication is included in the ``other'' time.
Clearly, almost all of the increase in the total time needed per simulation step can be attributed to the increase in the MPI communication time.
Furthermore at 64 GPUs a sharp increase in the MPI communication time is observed.
We presume that this can be attributed to the limited total network bandwidth of the single data rate InfiniBand installed in Lincoln.
Considering the relatively modest communication requirements of an MD simulation (at least for this simple Lennard-Jones system), this finding illustrates how important high throughput network connections are for GPU clusters.
In order to somewhat mitigate this problem, we have started to implement LAMMPS$_{\rm CUDA}$ modifications that will allow a partial overlap of force calculations and communication.
Preliminary results suggest that up to three quarters of the MPI communication time can be effectively hidden by that approach.

\begin{figure}[tb]
  \includegraphics[width=0.45\textwidth]{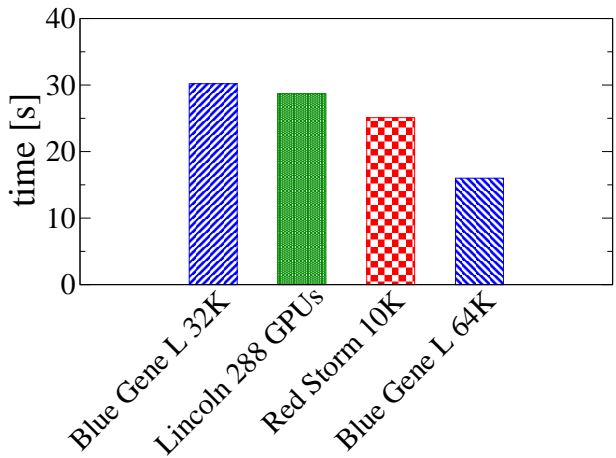}
  \caption{Loop time for 100 time-steps of a one billion particle Lennard-Jones system. On Lincoln, 288 GPUs were used. On the BlueGene/L system (with 32K processors and 64K processors) and RedStorm (with 10K processors), the regular CPU-based version of LAMMPS was used. System: LJ (see Appendix  \ref{sec:app_simulations}); Hardware: Lincoln (see Appendix  \ref{sec:app_hardware})}\label{fig:bench_billion}
\end{figure}

As a further example of what is possible with LAMMPS$_{\rm CUDA}$, we performed another large-scale simulation. 
Using 288 GPUs on Lincoln, we ran a one billion particle Lennard-Jones system (Density: 0.844, Cutoff: 2.5~$\sigma$). 
This simulation requires about 1~TB of aggregate device memory.
To the best of our knowledge, this is the largest MD simulation run on GPUs to date.
In Figure \ref{fig:bench_billion} loop times for 100 time-steps are shown for Lincoln, Red Storm (a Cray XT3 machine with 10368 processors sited at Sandia National Laboratories), and BlueGene/L (a machine with 65536 processor sited at Lawrence Livermore National Laboratory). 
The data for the latter two machines was taken from the LAMMPS homepage (http://lammps.sandia.gov/bench.html).
Using 288 GPUs, Lincoln required 28.7~s to run this benchmark, landing between Red Storm using 10,000 processors (25.1~s) and the BlueGene/L machine using 32K processors (30.2~s).

\section {Conclusion}

In this paper we have presented our own implementation of a general purpose GPU-MD code that we call LAMMPS$_{\rm CUDA}$.
This code already supports 26 different force field types.
We discussed multiple approaches for performing pair force calculations and concluded that an adaptive neighbor-list-based approach yields the best results.
Specifically, we have shown that the cell list approach is generally slower.
If running on a quad-core workstation with a single GPU, users can expect a 5x to 14x reduction in time-to-solution by harnessing the GPU, depending on the simulated system class (i.e. biomolecular, polymeric, granular, metallic, semiconductor).
With a strong focus on scalability, LAMMPS$_{\rm CUDA}$ can efficiently use the upcoming generation of GPU-based hybrid clusters, such as Tianhe-1A, Nebulae and Tsubame 2.0 (the first, third, and fourth fastest supercomputers on the November 2010 Top500 list).
By performing scaling benchmarks on up to 256 GPUs, LAMMPS$_{\rm CUDA}$ was shown to achieve general speed-ups of 20x to 60x using the latest generation of C1060s versus modern CPU cores, again depending on the simulated system class.
These numbers imply that using LAMMPS$_{\rm CUDA}$ on a 32 node system with 4 GPUs per node can achieve the same overall speed as the original CPU version of LAMMPS on a conventional CPU-based cluster with 1024 nodes.

\section {Acknowledgments}
This work was partially supported by the National Center for Supercomputing Applications by providing access to the Lincoln GPU cluster.
Sandia National Laboratories is a multi-program laboratory managed and operated by Sandia Corporation, a wholly owned subsidiary of Lockheed Martin Corporation, for the U.S. Department of Energy�s National Nuclear Security Administration under contract DE-AC04-94AL85000.

\appendix

\section{Benchmark simulations}
\label{sec:app_simulations}
\begin{itemize}
 \item \textbf{LJ}.
Potential: Lennard-Jones (lj/cut), Cutoff: 2.5~$\sigma_0$, Density: 0.84~$\sigma_0^{-3}$, Temperature: 1.6.
 \item \textbf{Silicate (cutoff)}.
Potential: Buckingham + Coulomb (buck/coul/cut), Cutoff: 15.0~\AA, Density: 0.09~\AA$^{-3}$,  Temperature: 600~K.
 \item \textbf{Silicate}.
Potential: Buckingham + Coulomb (buck/coul/long), Atoms: 11,664, Cutoff: 10.0 \AA, Density: 0.09~\AA$^{-3}$, Long range coulomb solver: PPPM (Precision: 2.4e-6),  Temperature: 600~K.
 \item \textbf{EAM}.
Potential: Embedded atom method (EAM)\cite{eam-1,eam-2}, Atoms: 256,000, Cutoff: 4.95~\AA, Density: 0.0847~\AA$^{-3}$, Temperature: 800~K.
 \item \textbf{Coarse Grained}.
Potential: Coarse grained systems (cg-cmm)\cite{cgcmm-1,cgcmm-2}, Atoms: 160,560, Cutoff: 15.0~\AA, Temperature: 300~K.
 \item \textbf{Rhodopsin}.
Potential: CHARMM force field + Coulomb (lj/charmm/coul/long)\cite{charmm}, Atoms: 32,000, Cutoff: 10.0~\AA, Long range coulomb solver: PPPM (Precision: 1e-7), Temperature: 300~K.
 \item \textbf{Granular}.
Potential: Granular force field (gran/hooke)\cite{gran-1,gran-2,gran-3}, Atoms: 1,152,000, Density: 1.07~$\sigma_0^{-3}$, Temperature: 19.
\end{itemize}


\section{Hardware systems} 
\label{sec:app_hardware}
\begin{small}
\begin{itemize}
\item Workstation Server A (WS$_A$)
  \subitem Intel Q9550 @ 2.8GHz
  \subitem 8GB DDR2 RAM @ 800 MHz
  \subitem Mainboard: EVGA 780i 3xPCIe2.0 16x
  \subitem 2 x NVIDIA GTX 280
  \subitem CentOS 5.4

\item Workstation Server B (WS$_B$)
  \subitem Intel i7 950 @ 3.0GHz
  \subitem 24GB DDR3 RAM @ 1066 MHz
  \subitem Mainboard: Asus P6X58D-E 3xPCIe2.0 16x
  \subitem 2 x NVIDIA GTX 470
  \subitem CentOS 5.5

\item GPU Cluster (CL)
  \subitem 2 x Intel X5550 @ 2.66GHz
  \subitem 48GB DDR3 RAM @ 1066 MHz
  \subitem Mainboard: Supermicro X8DTG-QF R 1.0a 4xPCIe2.0 16x
  \subitem 4 x NVIDIA Tesla C1060
  \subitem Scientific Linux 5.4

\item NCSA Lincoln (Lincoln)
  \subitem 192 Nodes 
  \subitem 2 x Intel X5300 @ 2.33GHz
  \subitem 16GB DDR2 RAM 
  \subitem 2 x NVIDIA Tesla C1060 on one PCIe2.0 8x (two nodes share one S1070)
  \subitem SDR Infiniband
  \subitem Red Hat Enterprise 4.8
\end{itemize}
\end{small}

\section{Calculations for comparing neighbor and cell lists} 
\label{sec:calc_comp}
In this section, we make use of the defintions from \ref{sec:comp_neigh_cell}, e.g. $t$ is the time for processing a single interaction and the fraction $\gamma \in [0, 1]$ of the time is assumed to be spent on memory accesses, i.e.:
\begin{equation}
 t = \underbrace{t \cdot \gamma}_{\text{memory read}} \;+\, \underbrace{t \cdot (1 - \gamma)}_{\text{actual calculation}}
\end{equation}
While the actual force calculation for one interaction is the same for both approaches, the time for memory access varies: it is assumed to be a factor $\alpha$ faster for the cell list approach, due to the \textit{coalesced accesses}.

The cell list will read $14$ neighbor cells and thus $N_{\text{neigh}} \propto 14 c^3$, while the neighbor list method will read $N_{\text{cell}} \propto \frac{4}{3} \pi c^3$ atoms. We assume a homogeneous density and thus the same proportionality factor $\varrho$ for both $N$, i.e. ($T = t \cdot N$):
\begin{eqnarray}
 T_{\text{neigh}} &=& \varrho \cdot \frac{4}{3} \pi c^3 \cdot t \cdot (\;\; \gamma \,\, + 1 - \gamma) \nonumber  \\
 T_{\text{cell}} &=& \varrho \cdot 14 c^3 \; \cdot t \cdot \left(\frac{\gamma}{\alpha} + 1 - \gamma\right) \;\;.
\end{eqnarray}

These formulas yield two interesting limiting cases while requiring that both approaches be equally fast ($T_{\text{neigh}} = T_{\text{cell}}$):
\begin{eqnarray}
 \lim_{\gamma \rightarrow 1} \left(\frac{1}{\gamma} \cdot \left(\frac{4 \pi}{42} - 1\right) + 1\right)^{-1} = \frac{42}{4 \pi} \approx 3.34 \\
 \lim_{\alpha \rightarrow \infty} \frac{\left(1 - \frac{4 \pi}{42}\right)}{\left(1 - \frac{1}{\alpha}\right)} = 1 - \frac{4 \pi}{42} \approx 0.70
\end{eqnarray}
In other words,
\begin{enumerate}[(i)]
 \item If {\it all} of the pair force time is used for memory accesses ($\gamma = 1$), then $\alpha$ has to be at least 3.3.
 \item If the memory accesses of the cell list approach take no time at all ($\alpha=\infty$), then $\gamma$ must still be $\approx 70$\%.
\end{enumerate}

\end{document}